\documentclass[pra,aps,twocolumn,showpacs,epsfig]{revtex4}
\usepackage{color}
\usepackage{graphicx,amssymb,epsfig,epsf,amsmath}

\begin{document}

\title{Dynamical features of Shannon information entropy of bosonic cloud in a tight trap}

\author{
Sudip Kumar Haldar$^{1}$\footnote{sudip\_cu@yahoo.co.in}, and Barnali Chakrabarti$^{2,3}$
} 

\affiliation{
$^{1}$Department of Physics, Lady Brabourne College, P1/2 Surawardi Avenue, Kolkata 700017, India\\
$^{2}$Department of Physics, Kalyani University, Kalyani, Nadia, West Bengal, India, Pin 741235\\
$^{3}$Instituto de Fisica, Universidade of S\~ao Paulo, CP 66318, 05315-970, S\~ao Paulo,Brazil
}

\begin{abstract}
We calculate Shannon information entropy of trapped interacting bosons in both the position and momentum spaces, $S_r$ and $S_k$ respectively. The total entropy maintains the fuctional form $S=a + b \ln N$ for repulsive bosons. At the noninteracting limit the lower bound of entropic uncertainty relation is also satisfied whereas the diverging behavior of $S_r$ and $S_k$ at the critical point of collapse for attractive condensate accurately calculates the stability factor. Next we study the dynamics of Shannon information entropy
 with varying interparticle potential. We numerically solve the time-dependent 
Gross-Pitaevskii equation  and study the influence of increasing 
nonlinearity in the dynamics of entropy uncertainty relation (EUR). 
We observe that for small nonlinearity the dynamics is regular. 
With increase in nonlinearity although Shannon entropy shows large 
variation in amplitude of the oscillation, the EUR is maintained throughout 
time for all cases and it confirms its generality. We also study the dynamics in a very tight trap when the condensate becomes highly correlated and strongly inhomogeneous. Time evolution of total entropy exhibits aperiodic and fluctuating nature in very tight trap. We also calculate 
Landsberg's order parameter for various interaction strengths which supports
earlier observation that entropy and order are decoupled.
\end{abstract}

\maketitle
\section{Introduction}

Statistical correlation and Shannon information entropies in both the position and momentum spaces are the key quantities in the understanding of quantum mechanical systems which measure information regarding localisation of the position and momentum distribution. An important step in this direction is the entropic uncertainty relation (EUR). Using the position and momentum space entropies ($S_r$ and $S_k$ respectively), Bialynicki-Birula and Mycielski (BBM) derived a stronger version of the Heisenberg uncertainty relation~\cite{I}. For a three-dimensional system the total entropic sum has the form 
\begin{equation}
S=S_r+S_k \ge 3(1+\ \ln{ (\pi)}) \simeq 6.434  .
\end{equation}
This means that the conjugate position and momentum space information entropies $S_r$ and $S_k$ maintain an inverse relationship with each other. It signifies that when a system is strongly localised in position space, the corresponding information entropy and uncertainty in position space decrease. The corresponding momentum distribution becomes delocalised, i.e. the information entropy and uncertainty in momentum space increase. It is important to note that BBM inequality~(Eq.(1)) presents a lower bound and the equality is maintained for a gaussian wave function.

In the past, work has been devoted to the study of the Shannon entropies in various quantum mechanical systems like nuclei, atomic clusters, fermionic and bosonic systems~\cite{Ohya,S,E,C,CP,Ch,Tutul}. The universal property of total entropy $S$ for the total density distribution in different quantum systems is proposed which takes the form~\cite{S}
\begin{equation}
S=a+b \ \ln{N} ;
\end{equation} 
$a$ and $b$ are the parameters which depend on the given systems. Information theory in a system of correlated bosons in a trap and various features of information entropy of different atomic systems are well studied~\cite{Massen, Sagar, Sen, Panos, Katriel}. Entanglement which is a central issue in quantum information theory has also been studied in different context~\cite{Abliz}. Shannon entropy has also been interpreted as a measure of electron correlation which is directly related to the strength of the interaction between the electrons~\cite{Robin}. In our present study we consider interacting trapped bosons in external confinement at zero temperature. The gas is considered to be extremely dilute as the inter-atomic separation is quite large compared to the range of interatomic interaction and the fundamental features of such trapped bosons are accurately described by a single parameter $a_{sc}$, the $s$-wave scattering length. This is quite different from the dipolar bosons for which interaction potential includes long-range anisotropic dipole-dipole interaction (in addition to the short-range $\delta$ interaction) which has significant effect on the phase transition of the dipolar BEC in an optical lattice~\cite{Xie}. The choice of our system is quite similar to that of earlier studies~\cite{Massen}, however the motivation of our present work is quite different, as follows. First: we are interested in the study of dynamics of Shannon entropies and the EUR in a fixed confinement but with a varying effective interaction $Na_{sc}$. For weakly interacting large number of bosons in the external confinement, the ground state properties are quite accurately explained by the nonlinear equation, known as time-independent Gross-Pitaevskii (GP) equation
\begin{equation}
\Big[-\frac{\hbar^2}{2m}\bigtriangledown^2
+\frac{1}{2}m\omega^2r^2
+g|\psi(\vec{r})|^2 \Big]\psi(\vec{r}) = \mu \psi(\vec{r}) 
\end{equation}
where $g$ is the interaction strength parameter given by $g=\frac{4\pi \hbar^{2} a_{sc} N}{m}$. $N$ is the number of atoms, $m$ is the atomic mass, $a_{sc}$ is the dimer scattering length, and $\omega$ is the trap frequency. In the earlier works in this direction, Eq.~(3) has been solved and the EUR is tested for various number of atoms~\cite{Massen}. However in the earlier study the effect of nonlinearity in the dynamical feature of information entropies is not focussed. For this one should solve the time-dependent GP equation and using the dynamics of the many-boson wave function one can directly observe the effect of nonlinearity in the time evolution of information entropy and the EUR. Although the nonlinear effect in the time evolution of the condensate has been studied earlier~\cite{Ruprecht, Berman, Pierre, Villain}, the dynamics of information entropy and the validity of the EUR with evolution in time is a stronger tool to manifest the effect of nonlinearity and it deserves special interest. Second: we study the dynamics of information entropies in very tight trap. In the present day experiment, by controlling the external magnetic field, one can virtually manipulate the external trap size. In very tight trap, the atoms will be highly correlated and strongly inhomogeneous. The calculation of Shannon entropies will be very informative to describe such a highly localised and strongly inhomogeneous system. The study of dynamical features of information entropies and the dynamics of EUR will give rich physics.

This paper is organised as follows. Sec. II deals with the brief review of time dependent GP equation and calculation of dynamics of Shannon entropies. Sec. III deals with our results. Here we analyze the dynamics of EUR in different trap sizes for various interaction strength parameters~$g$. We also present our calculated values of Landsberg's order parameter $\Omega$ for various interaction strength parameters~$g$ at time $t=0$. Sec. IV concludes with summary.

\section{Dynamics of the Shannon information entropy}

We start with the zero temperature condensate in the harmonic trap 
described by the time dependent Gross-Pitaevskii equation as 
\begin{equation}
 i\hbar \frac{\partial \psi({\vec{r}},t)}{\partial t} =
\Big[-\frac{\hbar^2}{2m}\bigtriangledown^2
+\frac{1}{2}m\omega^2r^2
+g|\psi({\vec{r}},t)|^2 \Big]\psi({\vec{r}},t) ,
\end{equation}
To solve Eq. (4) we start with the analytic and normalized 
ground state solution in the absence of the nonlinear term and 
propagate the GP equation with time. The numerical integration 
of the time-dependent GP is obtained by using a finite-difference 
Crank-Nicolson algorithm with a split operator technique~\cite{cerb,sala,gmaz}.
We have verified that the numerical solution of GP equation is
consistent with various sets of radial grids and time steps. 
In the time-dependent solution although the probability 
is time independent, i.e stationary in time, 
the condensate wave function 
has time dependent phase factor as $\exp(-i\mu t)$, where $\mu$ is the 
energy in oscillator unit. Thus the effect of perturbation 
comes through the nonlinear term which involves the time dependent wave
function in the previous time step and allows us to monitor
the condensate motion with time. 
\\
Next we calculate the time-dependent momentum
space wave function $\phi(\vec{k},t)$ by taking Fourier transformation 
of the time-dependent position space wave function $\psi(\vec{r},t)$ at each time step
 as 
\begin{equation}
\phi(\vec{k},t) = \frac{1}{(2 \pi)^{3/2}}\displaystyle\int{\psi(\vec{r},t) \ 
e^{-i\vec{k}\cdot\vec{r}}d^{3}\vec{r}} .
\end{equation}
Here $\psi(\vec{r},t)$ is the normalized ground state solution of Eq.~(4) and is given by $\psi(\vec{r},t) = \frac{1}{\sqrt{4 \pi}} \psi(r,t)$, where $\psi(r,t)$ is the radial part of the wave function. The angular part of the ground state solution is a constant and is taken care by the constant pre-factor along with the normalization. Now due to the presence of spherically  symmetric trap, all directions are equivalent and we set the polar axis of the spherical polar coordinate along the direction of $\vec{k}$. Then performing the integrals over the angular coordinates first, the $3D$ integral of Eq.~(5) simplifies to an integration over the radial coordinate only.
\begin{equation}
\displaystyle\int{\psi(\vec{r},t) \ 
e^{-i\vec{k}\cdot\vec{r}}d^{3}\vec{r}} = \frac{2 \sqrt{\pi}}{k}\displaystyle\int {r \psi(r,t) sin (kr) dr} .
\end{equation} 
Here $k$ is the magnitude of $\vec{k}$. We perform this integral over $r$ by using a 32-point gaussian quadrature. We truncate the sum at $r=r_{max}$ which is detrmined from the asymptotic behavior and normalization of $\psi(r,t)$.

For a three-dimensional system the information 
entropy in the position space is calculated from the density 
distribution $\rho(\vec{r},t)$ = $|\psi(\vec{r},t)|^{2}$ as 
\begin{equation}
S_{r}(t) = -\int {\rho(\vec{r},t)} \ \ln{\big(\rho(\vec{r},t)\big)} 
\ d^{3}\vec{r} \; . 
\end{equation}
The corresponding information entropy in the momentum space is determined 
from the 
momentum distribution $n(\vec{k},t)$= $|\phi(\vec{k},t)|^{2}$  as
\begin{equation}
S_{k}(t) = -\int {n(\vec{k},t)} \ \ln{\big( n(\vec{k},t) \big)} \ d^{3}\vec{k} 
\; .
\end{equation}
If $\rho(\vec{r},t)$ and $n(\vec{k},t)$ are normalized to unity, the joint 
entropy at time $t$ is defined as $S(t) = S_{r}(t) + S_{k}(t)$. At $t=0$ as there is no effect of nonlinear terms, $S(t=0)$ should obey the existing entropic uncertainity relation (EUR)~\cite{I} given as
\begin{equation}
\begin{array}{cl}
S(t=0) &= S_{r}(t=0) + S_{k}(t=0)\\
 & \ge  S_{min}=3 (1+ \ln{(\pi)}) \simeq 6.434 \\ \;
\end{array} .
\end{equation}
It is already pointed out in the study of different 
quantum systems that the EUR is stronger than the Heisenberg's 
uncertainity relation due to the following reasons~\cite{CP}.
First: Heisenberg's relation can be derived from the EUR but its 
reverse is not true. Second: Heisenberg's relation depends on the 
state of the system but the EUR does not~\cite{CP}. 

Although Eq. (5) - (9) have been used earlier in the study of interacting bosonic systems, however it needs additional discussion regarding the validity of the expression for a condensate having finite size. Although Eq.~(7) and Eq.~(8) are apparently valid for single particle only, however its extension and use for the condensate is equally valid. Above the critical temperature $T_c$ although there is a contribution coming from non-condensed atoms, but at the critical temperature $T_c$ ($\sim$ nk) there is a macroscopic occupation of the bosons. At such low temperature, the de Broglie wave-lengths of the neighbouring atoms overlap, the individual atoms lose their quantum identity. Thus the whole bosonic cloud is treated as a single quantum stuff which is well described by the zero-temperature GP equation. This is also in perfect agreement with the framework of generalized Bogoliubov prescription, where the condensate wave function is the  expectation value of the field operator. Thus the fluctuation in field operator is neglected. This is quite justified as the depletion of the condensate at the the temperature $\sim$ nk is negligible. Thus Eq.~(7) and Eq.~(8) basically determines the information entropy associated with the whole condensate and not of individual bosons. 

\section{Results}

We consider the condensate of $^{87}$Rb atoms with $a_{sc}$ = 100 $a_0$ and initially choose the trap frequency $\omega$ as 77.87 Hz which mimics the JILA experiment~\cite{Anderson}. With these parameters we solve numerically Eq.~(4) for various number of bosons in the trap.  We calculate the condensate wave function in the coordinate space with varying nonlinearity $g$. In this work we express all the quantities in oscillator unit (o.u.). In oscillator units (o.u.) all lengths are expressed in units of $a_{ho}$ ($= \sqrt{\frac{\hbar}{m \omega}}$) and all energies in units of $\hbar \omega$, $\omega$ being the trap frequency. Since the oscillator units (o.u.) are frequency dependent, for our present work we define them ({\it i.e.} $a_{ho}$ and $\hbar \omega$) with respect to the frequency $\omega_0$ (=77.87 Hz) of the JILA trap~\cite{Anderson}. First to check the consistency of our results with earlier published results~\cite{Massen,Sagar,Ruprecht}, we solve the time-independent GP equation (Eq.~(3)) with varying number of bosons $N$ (or equivalently with varying strength of interatomic interaction $g$) for pure harmonic trap. In Fig.~1(a) we plot the condensate wave functions $|\psi|$ for various number of atoms in the trap. For a fixed external trap the wave function $|\psi|$ gradually expands with increase in number of bosons as the strength of repulsive interaction $g$ (=$\frac{4\pi \hbar^{2} a_{sc} N}{m}$) increases with increase in $N$. 
\begin{figure}
  \begin{center}
    \begin{tabular}{cc}
      \resizebox{80mm}{!}{\includegraphics[angle=0]{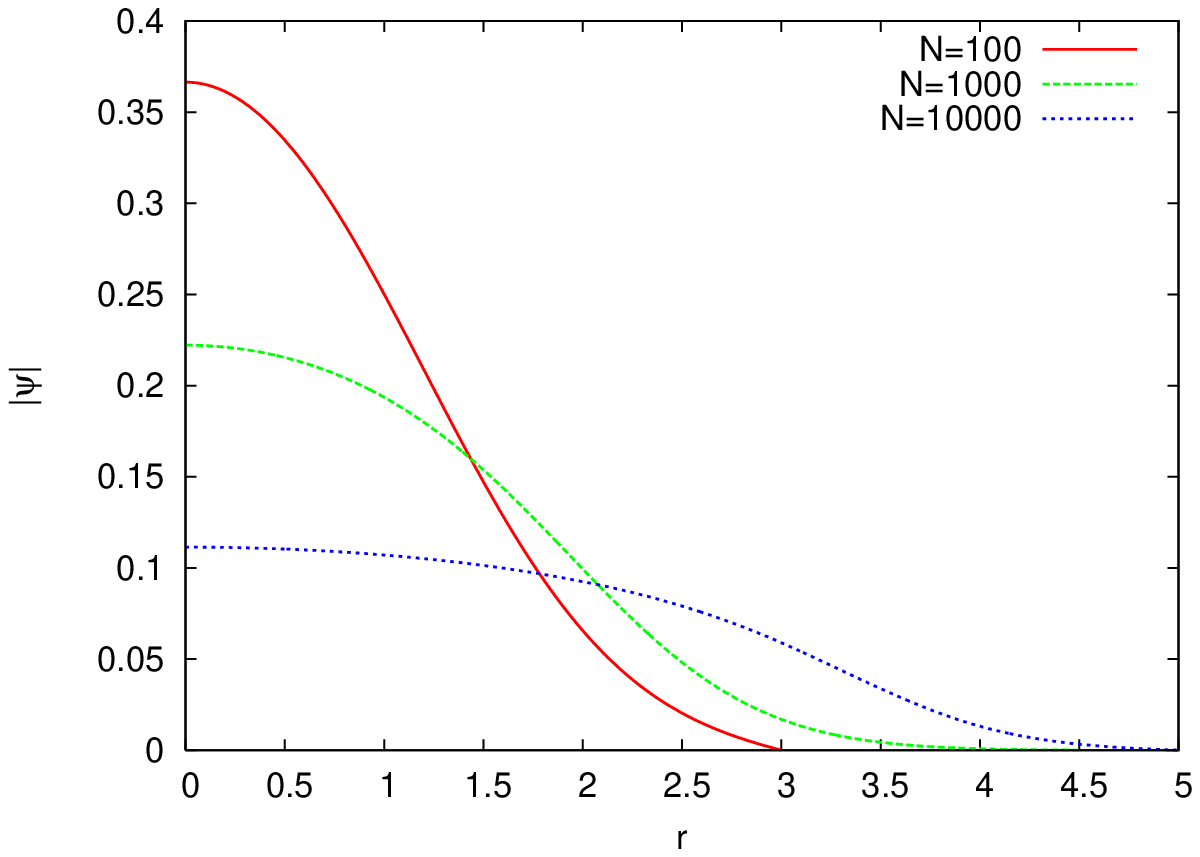}} &\\
         (a)   &\\
      \resizebox{80mm}{!}{\includegraphics[angle=0]{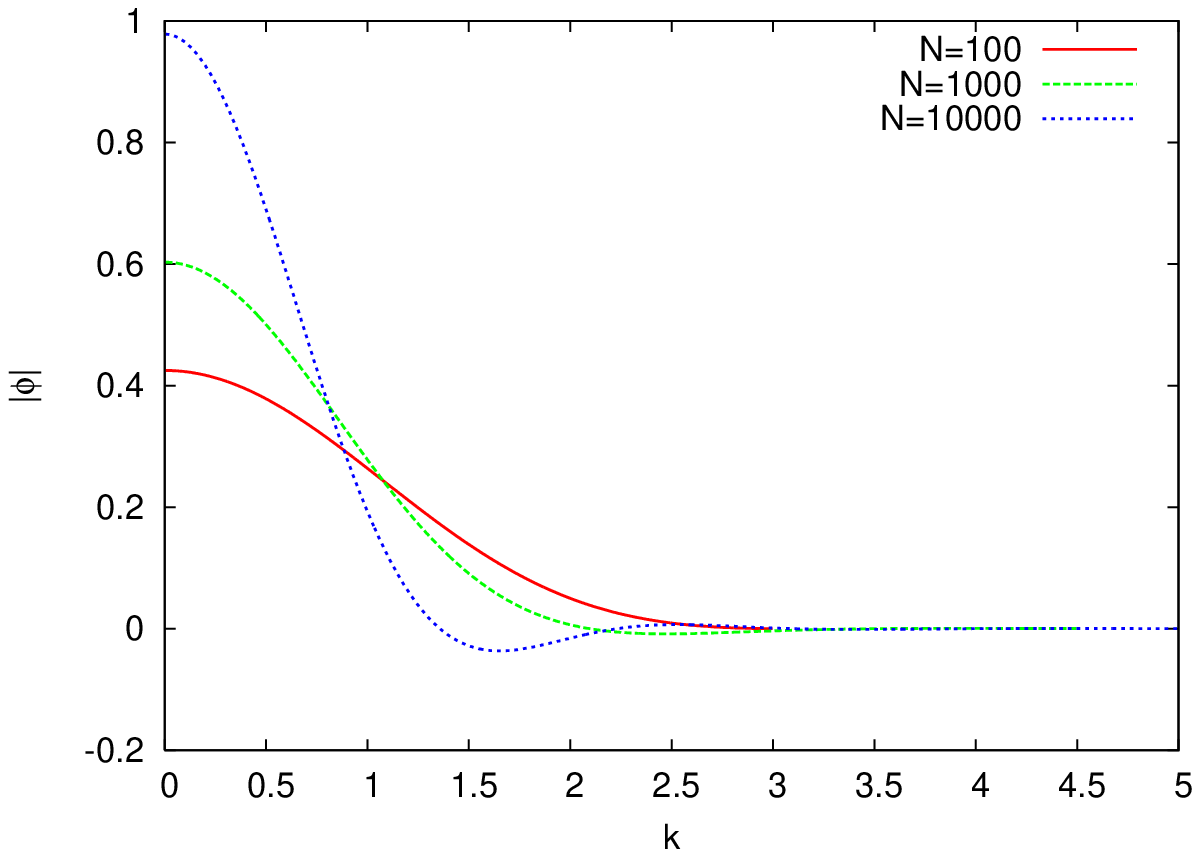}} & \\
           (b) & \\  
    \end{tabular}
  \end{center}

\caption{(color online) (a) Plot of the condensate wave function for repulsive BEC of $^{87}$Rb atoms in harmonic trap for various $N$. (b) Plot of the corresponding momentum space wave functions. }
\end{figure}
\begin{figure}
  \begin{center}
    \begin{tabular}{cc}
      \resizebox{80mm}{!}{\includegraphics[angle=0]{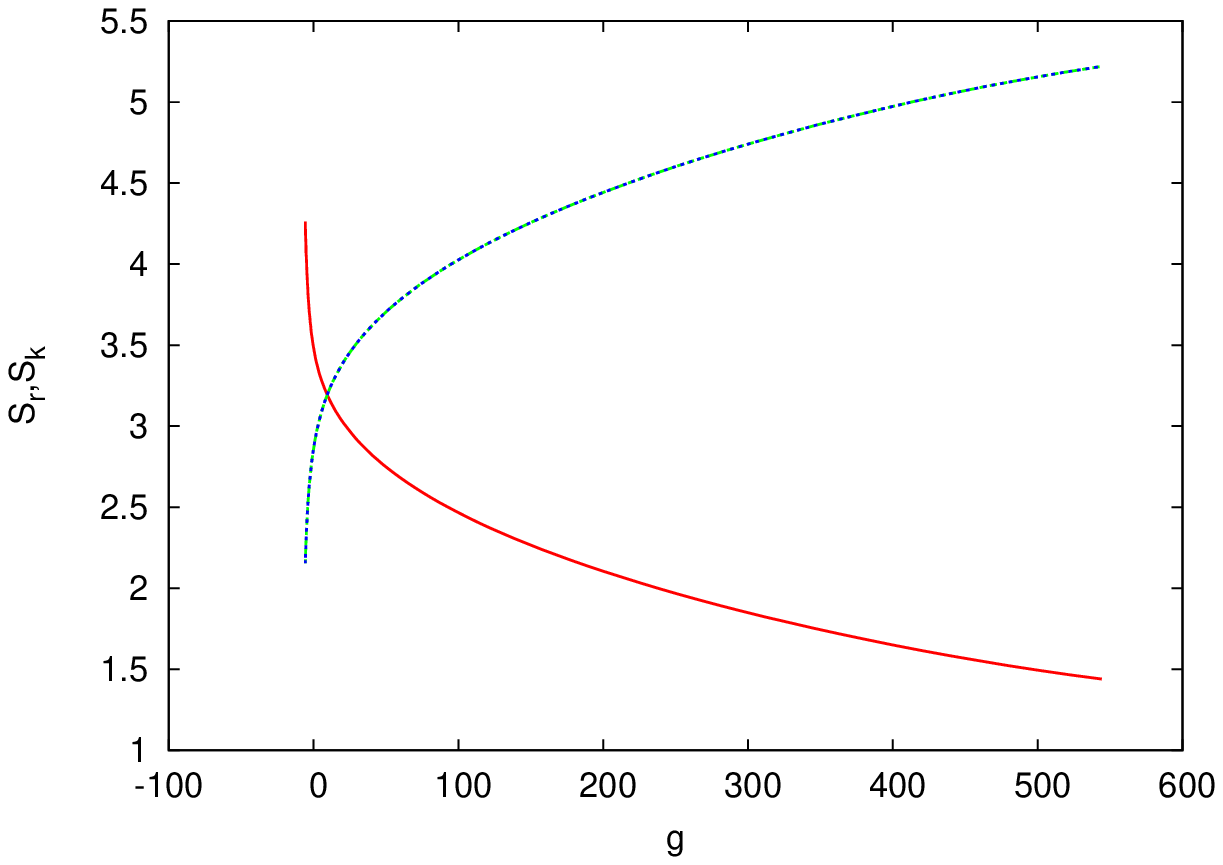}} &\\
         (a)   &\\
      \resizebox{78mm}{!}{\includegraphics[angle=0]{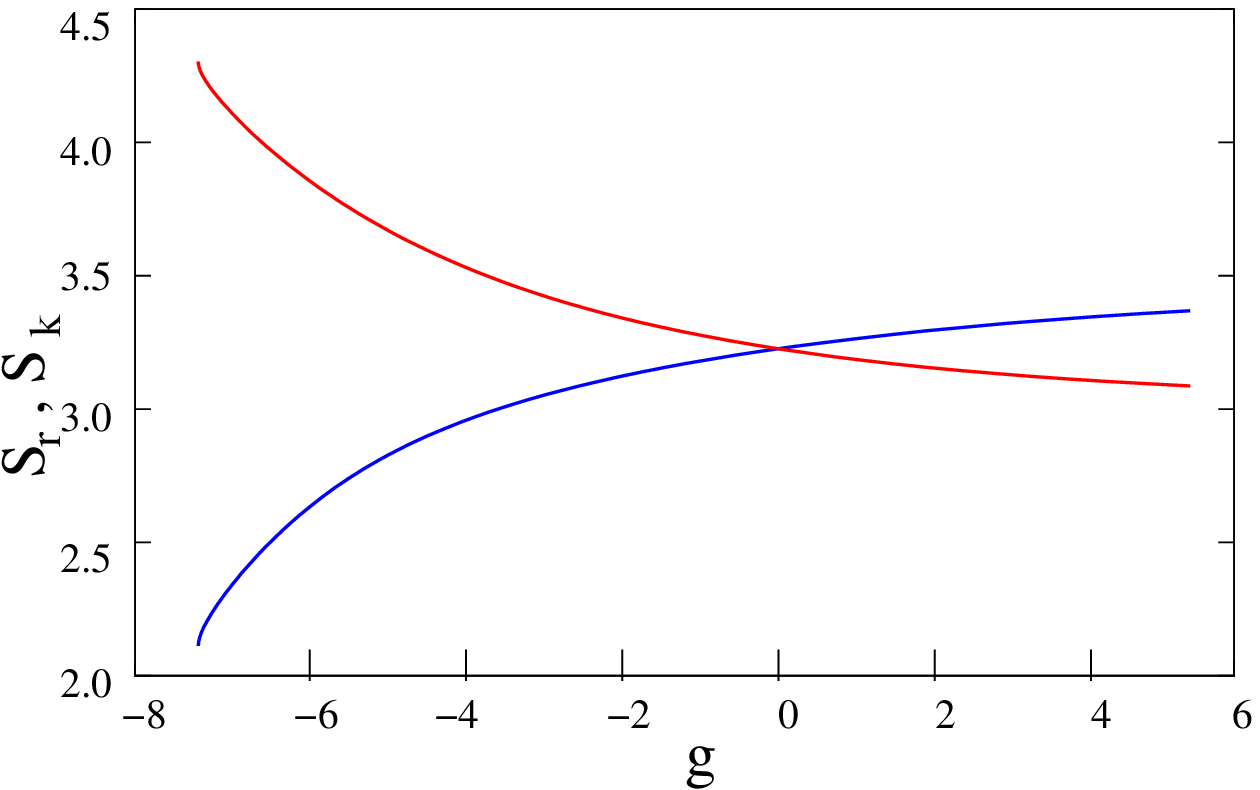}}& \\
           (b) & \\  
    \end{tabular}
  \end{center}
\caption{(color online) a) Plot of $S_r$ (blue dotted curve) and $S_k$ (red smooth curve) for various nonlinearity $g$; b) Enlarged view of the -ve $g$ part.}
\end{figure}
The momentum space wave function $|\phi|$ is calculated by Fourier transformation of the coordinate space wave function $|\psi|$. In Fig.~1(b) we plot condensate wave functions in momentum space $|\phi|$ for the same values of $N$ as in Fig.~1(a). We see the expected reciprocal behavior between the position and momentum space wave functions in accordance with the Heisenberg's uncertainty principle. Initially for small $N$, the effect of nonlinear interaction is quite small and the condensate wave function is very close to gaussian. The momentum space condensate wave function is also close to gaussian. As we increase $N$, the net nonlinear interaction increases, the coordinate space condensate wave function $|\psi|$ spreads out. The corresponding momentum space wave function $|\phi|$ squeezes accordingly. For large $N$ (=10,000) $|\psi|$ is quite diffuse in coordinate space and $|\phi|$ is sharply peaked in momentum space and has a small kink which is due to the presence of finite size trap. Similar kink is also observed in earlier studies~\cite{Giorgini}. This also nicely demonstrates how the presence of the harmonic trap modifies the momentum distribution. Next we calculate the information entropy $S_r$ as $S_{r} = -\int {\rho(\vec{r})} \ \ln{\big(\rho(\vec{r})\big)} \ d^{3}\vec{r} $ and the conjugate information entropy in momentum space $S_k$ as $S_{k} = -\int {n(\vec{k})} \ \ln{\big( n(\vec{k}) \big)} \ d^{3}\vec{k}$. In Fig.~2 we plot $S_r$ and $S_k$ as a function of non linear interaction $g$. With increase in particle number $N$, the strength of the net repulsive interaction $g$ increases, the coordinate space wave function $|\psi(\vec{r})|$ delocalises and consequently the position space entropy $S_r$ gradually increases. It signifies that the associated uncertainty in the coordinate space increases. The corresponding momentum space wave function $|\phi(\vec{k})|$ is localised and the momentum space entropy $S_k$ gradually decreases with increase in particle number $N$ (i.e. $g$). It indicates that the associated uncertainty in momentum space decreases when net effective repulsive interaction increases. This inverse behavior between the position and momentum space entropies in Fig.~2 is interpreted as a consequence of the EUR and it also reflects the reciprocal behavior of wave functions in position and momentum spaces which is the consequence of Heisenberg uncertainty relation. At $g=0$, we have the noninteracting inhomogeneous ideal Bose gas. The important consequence of the inhomogeneity of the system is that BEC appears not only in momentum space but also in coordinate space. In the absence of nonlinear interatomic interaction, the condensate wave function becomes perfectly gaussian both in coordinate and momentum spaces. Consequently the Shannon information entropy in position space $S_r$ and that in momentum space $S_k$ become equal to each other and we find $S_r=S_k=3.217$. Thus the total information entropy $S=S_r+S_k$ becomes exactly equal to 6.434 which is the lower bound of EUR. It is to be noted that the gaussian wave functions correspond to the coherent state with the associated Heisenberg's uncertainty product being minimum. For attractive potentials ($g<0$), the behavior is opposite to that observed for repulsive potential. It is well known that in case of attractive BEC, the condensate tends to increase its density in the centre of the trap in order to lower its interaction energy. This tendency is balanced by the zero-point kinetic energy which can stabilize the system. However for large number of bosons in the trap, the central density grows too much and the kinetic energy can not balance it anymore. The system thus collapses when the number of particles in the condensate exceeds a  critical number $N_{cr}$. In the mean-field GP equation the critical number is determined in the following way. For attractive interaction and $N<N_{cr}$, the condensate is always metastable and the energy functional has a local minimum. When $N$ increases, the depth of local minimum decraeses and exactly at $N=N_{cr}$, the minimum vanishes, and the GP equation has no solution. The calculated value of stability factor for a spherical trap is $\frac{N_{cr}|a_{sc}|}{a_{ho}}=0.575$~\cite{Ruprecht}. In our present study  we observe that with increase in effective negative interaction $S_r$ sharply falls which indicates that the condensate wave function in coordinate space $|\psi|$ squeezes in the centre of the trap. Whereas the corresponding momentum wave function $|\phi|$ spreads out and $S_k$ sharply increases as expected. Finally at $g=-7.23$, both $S_r$ and $S_k$ diverge. This corresponds to the stability factor $k_{cr}=\frac{N_{cr}|a_{sc}|}{a_{ho}}=0.575$ which is same as that obtained from GP theory. Thus we can correlate the collapse of the attractive BEC with the simultaneous divergence of $S_r$ and $S_k$. The divergence in $S_r$ nicely shows how the probability distribution in the coordinate space becomes sharply localised at the centre of the trap, like a $\delta$-function. Thus our present study gives us more physical insight in the description of the collapse for attractive forces.

\begin{figure}[hbpt]
\vspace{-10pt}
\centerline{
\hspace{-3.3mm}
\rotatebox{0}{\epsfxsize=8cm\epsfbox{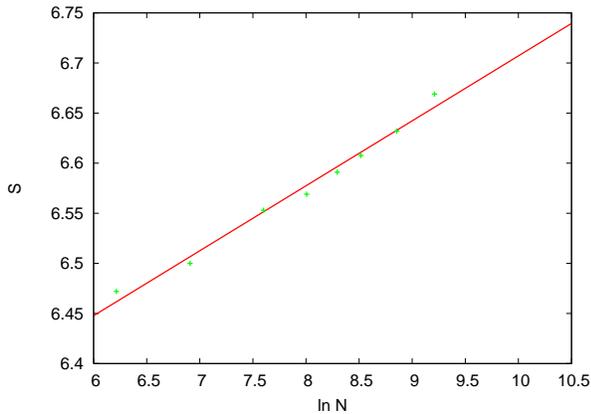}}}

\caption{(color online) Plot of the total information entropy $S$ vs $\ln {N}$. Dots present the numerically calculated data and the straight line is the plot of Eq.~(2) with calculated values of $a$ and $b$.}
\end{figure}

It is an established fact that for repulsive bosons in the trap there is no limitation about the maximum number of atoms as the condensate is always stable for any number of atoms. Thus if we continue to increase $N$, $S_r$ will increase and $S_k$ will decrease accordingly as described earlier. However due to the inhomogeneous nature of the system the increase in $S_r$ is not completely balanced by the corresponding decrease in $S_k$. Consequently the total entropy $S$ will increase as shown in the Fig.~3 where we plot total entropy $S$ as a function of $\ln {N}$. We find the plot of $S$ vs $\ln {N}$ to be a straight line of the form $S=a + b \ln {N} $ and it conforms to the earlier observation by Massen {\it et. al.}~\cite{Massen}. Our calculated values of $a=6.059$ and $b=0.065$ are also in good agreement with earlier findings~\cite{Massen}.   

Next, we are interested in study of the dynamical features of entropies in both the harmonic and tight trap. We first calculate the time dependenct local density in coordinate 
space $\rho(\vec{r},t)$ and in momentum space $n(\vec{k},t)$ from the 
corresponding condensate wave functions $|\psi(\vec{r},t)|$ and  
$|\phi(\vec{k},t)|$ respectively. Then from 
the local density $\rho(\vec{r},t)$ = $|\psi(\vec{r},t)|^{2}$, we 
calculate $S_{r}(t)$ by utilizing Eq.~(6) and from $n(\vec{k},t)$ = $|\phi(\vec{k},t)|^{2}$ 
we determine $S_{k}(t)$ by Eq.~(7) and then the total entropy $S(t)$. We determine $S_{r}(t)$, $S_{k}(t)$ and 
$S(t)$ with increasing interaction strength $g$ to observe the effect of 
nonlinearity in the time evolution of the information entropies and the dynamics of EUR.

In pure harmonic trap the interacting bosons are inhomogeneous. Thus the increase in $S_r$ with particle number does not cancel exactly with the decrease in $S_k$ or vice versa. However the effect of inhomogeneity is quite small for very small number of atoms in the trap. At the same time due to the presence of nonlinear interaction the dynamical features of $S_r$ and $S_k$ may exhibit some nonlinear effect in time evolution. As the nonlinear effect increases with particle number, we expect to get strong nonlinear features even in pure harmonic trap when the effect of nonlinear interaction~$g$ is quite large. Thus depending on the total number of atoms in the external trap, the time evolution of $S_r$, $S_k$  and the EUR can highlight the point of nonlinearity as the Shannon entropies are the best measures of localisation or delocalisation of the distribution function. The results for $g=10$ are presented in Fig.~4. For $g$=10, the number of bosons is $N=184$. As the system is very dilute, for such weakly interacting system the corresponding distribution function is close to gaussian.Thus we predict linear, periodic, behavior in $S_r(t)$, $S_k(t)$ and $S(t)$. Total entropy $S(t)$ starts from $6.434$ at $t=0$ and then exhibits regular oscillation with very small amplitude. The increase in $S_r(t)$ and the corresponding decrease in $S_k(t)$ with time perfectly satisfies the physical meaning of the inequality Eq.~(8)~\cite{I}. We also note that the maxima in $S_r(t)$ corresponds to minima in $S_k(t)$ and vice versa. Thus the joint measures of the uncertainty clearly signifies that for few hundreds of bosons, the system is very close to linear. The small effect of nonlinearity  is smeared off by the external trap as the interaction energy is neglible compared to trap energy at zero temperature.
\begin{figure}[hbpt]
\vspace{-10pt}
\centerline{
\hspace{-3.3mm}
\rotatebox{0}{\epsfxsize=8cm\epsfbox{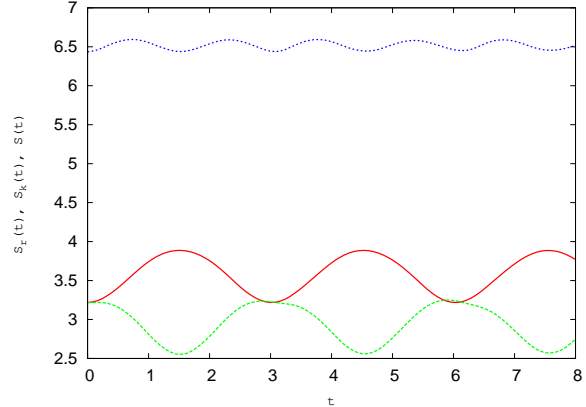}}}

\caption{(color online) Plot of $S_r(t)$ (red smooth curve), $S_k(t)$ (green dashed curve), and $S(t)$ (blue dotted curve) with time $t$ for $g=10$ and $\omega=1.0 \omega_0$ .}
\end{figure}

In atomic physics the total entropy 
maintains some rigorous inequalities which can be derived using 
the EUR. $S_{r}$ and $S_{k}$ are fundamentally 
related with the total kinetic energy T and the mean square radius 
$\langle r^2 \rangle$ of the system~\cite{Massen}. It can be shown~\cite{Gadre} that 
\begin{equation}
S_{min} \le S(t) \le S_{max}(t) \; , 
\end{equation}
where the lower limit of total entropy is 
the previously introduced constant and 
the upper limit of the total entropy is 
\begin{equation}
S_{max}(t)= 3
(1+\ln{(\pi)})+\frac{3}{2}\ln{\Big(\frac{8}{9}\langle r^2\rangle_t T(t)\Big)} \; ,  
\end{equation} 
where $\langle r^2 \rangle_t$ is the mean square radius at time $t$ 
and $T(t)$ is the kinetic energy of the system. In this paper 
we analyze the time evolution of the inequality by 
calculating  time evolution of  $\langle r^2 \rangle_t$ and $T(t)$ 
and $S_{max}(t)$ according 
to Eq.(10). We study how the above inequality Eq.~(9) is maintained with time.

We plot $S_{max}(t)$ with total $S(t)$ in Fig.~5. for $g=10$. 
The inequality is maintained with time nicely. So the system is 
very close to linear. 
\begin{figure}[hbpt]
\vspace{-10pt}
\centerline{
\hspace{-3.3mm}
\rotatebox{0}{\epsfxsize=8cm\epsfbox{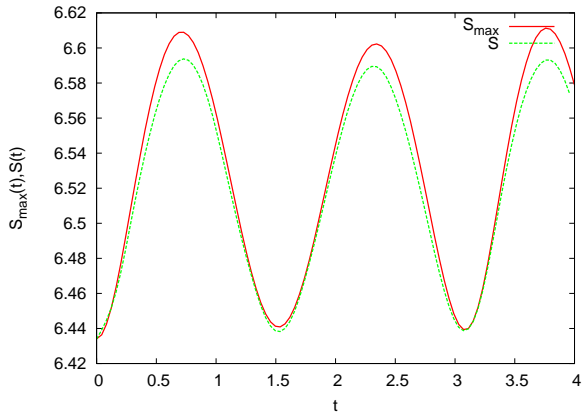}}}

\caption{(color online) Plot of $S(t)$ (red smooth curve) and $S_{max}(t)$ (green dashed curve) with time $t$ for $g=10$.}
\end{figure}

\begin{figure}[hbpt]
\vspace{-10pt}
\centerline{
\hspace{-3.3mm}
\rotatebox{0}{\epsfxsize=8cm\epsfbox{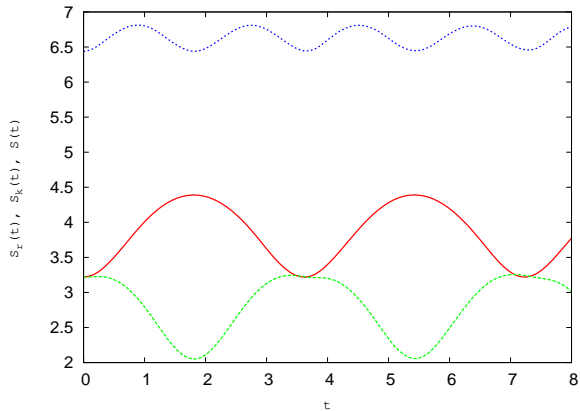}}}

\caption{(color online) Plot of $S_r(t)$ (red smooth curve), $S_k(t)$ (green dashed curve), and $S(t)$ (blue dotted curve) with time $t$ for $g=10$ in wide trap ($\omega= 0.85 \omega_0$).}
\end{figure}

\begin{figure}[hbpt]
\vspace{-10pt}
\centerline{
\hspace{-3.3mm}
\rotatebox{0}{\epsfxsize=8cm\epsfbox{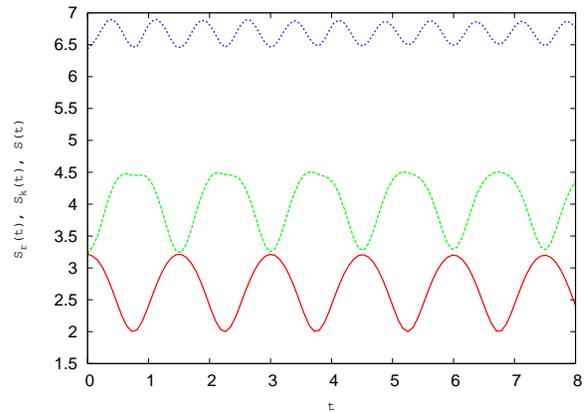}}}

\caption{(color online) Plot of $S_r(t)$ (red smooth curve), $S_k(t)$ (green dashed curve), and $S(t)$ (blue dotted curve) with time $t$ for $g=10$ in very tight trap ($\omega=2.0 \omega_0$).}
\end{figure}

Next we observe the effect of the change of trap size on the dynamics of $S_r$, $S_k$ and $S$ for few hundreds bosons ($g=10$). The trap size can be changed by varying the trap frequency $\omega$ (since the trap width $\approx a_{ho}$). Here we express various trap sizes by their corresponding trap frequncy $\omega$, measured as a multiple of the frequency $\omega_0$ of the JILA trap~\cite{Anderson}. The results with $\omega=0.85 \omega_0$ (wide trap) and $\omega=2.0\omega_0$ (very tight trap) are presented in Fig.~6 and Fig.~7 respectively. For $\omega=0.85 \omega_0$, the external trap spreads compared to pure harmonic trap. The corresponding condensate wave function in coordinate space spreads accordingly and the momentum space wave function is squeezed. It is reflected in Fig.~6 nicely. As $|\psi(\vec{r},t)|$ is diffuse in such a wide trap $S_r(t)$ increases and $S_k(t)$ decreases. For such small number of bosons in wide trap the system is less inhomogeneous and very close to linear. Thus increase in $S_r(t)$ and the corresponding decrease in $S_k(t)$ almost cancell each other and the total entropy makes a regular periodic oscillation with small amplitude. However the effect of larger trap size is clearly seen in evolution of total entropy $S(t)$. For $\omega=1.0\omega_0$ (pure harmonic trap), we observe five distinct maxima in the total time period ($t=0$ to $t=8$). Whereas for $\omega=0.85 \omega_0$, $S(t)$ exhibits four distinct maxima in the same period. The position of the first maximum in $S(t)$ occurs at $t=0.8$ for $\omega=1.0\omega_0$ wheras the same occurs in $S(t)$ at $t=0.96$ for $\omega=0.85 \omega_0$. Thus the wavelegth increases in larger trap size which is in perfect agreement with earlier observation that effect of correlation decreases in larger trap~\cite{Anindya}. 

The results for tight trap with $\omega=2.0\omega_0$ are presented in Fig.~7. In such a tight trap the system is strongly inhomogeneous but not strongly nonlinear as the number of atoms are just few hundreds. As now the condensate wave function is strongly localised in coordinate space, the corresponding uncertainty decreases and the $S_r$ also decreases sharply. Still as the system is close to linear due to such small number of atoms, $S_r(t)$ maintains regular periodic oscillation. Whereas we observe reciprocal behavior in $S_k(t)$  which increases and exhibits periodic oscillation, the total entropy $S(t)$ also shows regular periodic behavior but with many oscillations compared to $\omega=0.85 \omega_0$ and $\omega=1.0\omega_0$ cases. The first maximum now appears at $t=0.32$  which shows that wavelength of the oscillation becomes small and indicates high spatial coherence in tight trap~\cite{Anindya}.

To observe the effect of nonlinearity we do the same for large number of particles which corresponds to large nonlinear interaction. Results for $g=100$ are presented in Fig.~8 where we plot the evolution of total entropy $S(t)$ for $\omega=0.85 \omega_0$, $\omega=1.0\omega_0$ and $\omega=2.0\omega_0$. For pure harmonic trap with $N=1840$ ($g=100$), we observe that $S(t)$ starts with 6.434 at $t=0$ and then oscillates with large amplitude. Comparing with a similar figure for $N=184$, the amplitude of oscillation is not absolutely constant throughout the whole period. It shows that though the system is not nonlinear, however it goes away from equilibrium. This is expected as the increase in $S_r(t)$ does not cancel the decrease in $S_k(t)$ as in earlier case. Thus for $g=100$, the effect of inhomogenity is visible in the dynamics of $S(t)$. For $\omega=0.85 \omega_0$, we also   
\begin{figure}[hbpt]
  \begin{center}
    \begin{tabular}{cc}
      \resizebox{80mm}{!}{\includegraphics[angle=0]{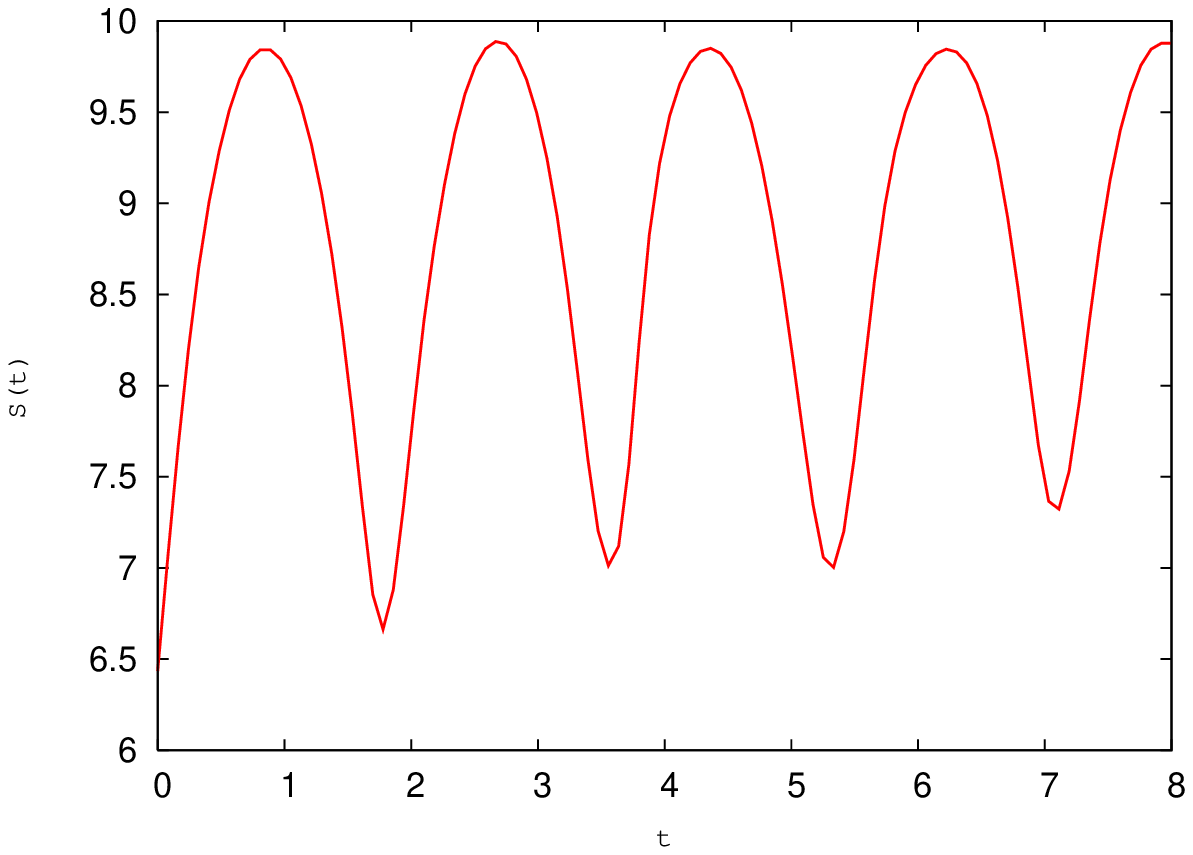}} &\\
         (a) $\omega =0.85$ &\\
      \resizebox{80mm}{!}{\includegraphics[angle=0]{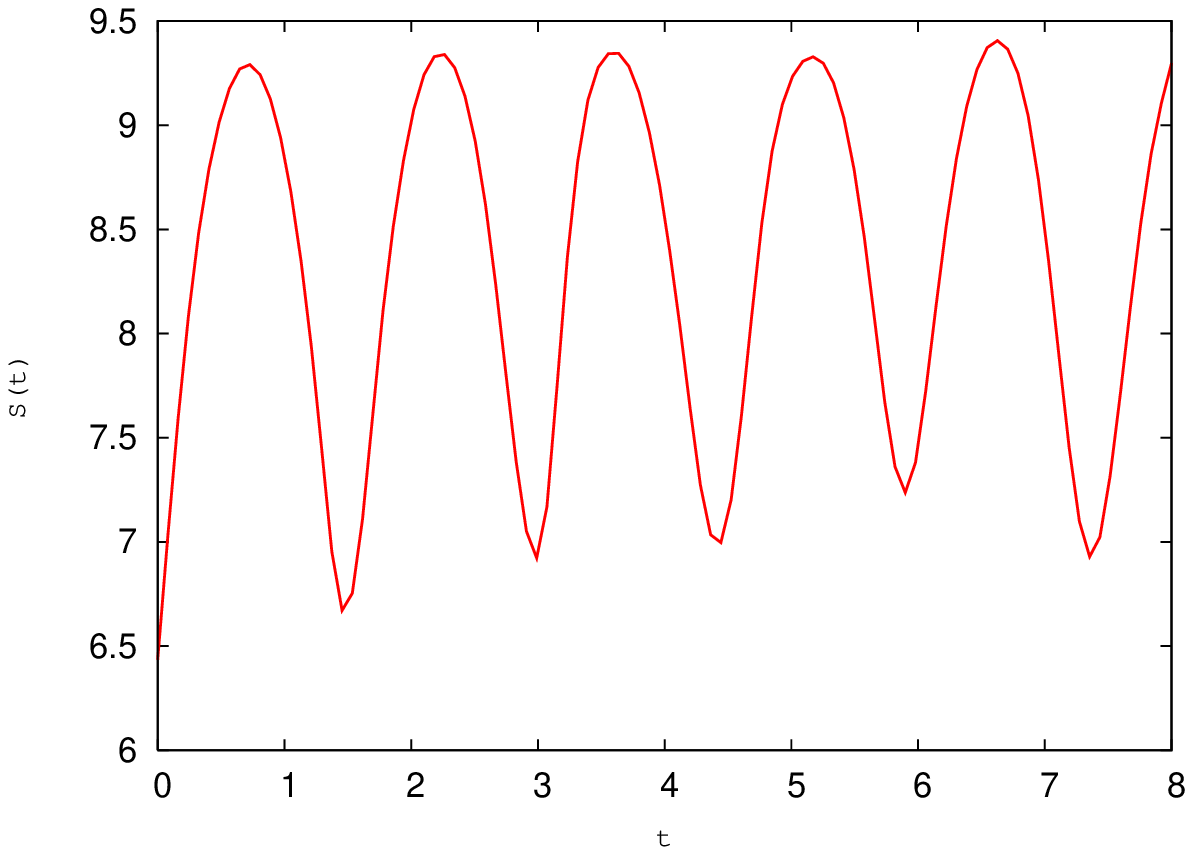}} & \\
         (b) $\omega =1.0$ &\\  
      \resizebox{80mm}{!}{\includegraphics[angle=0]{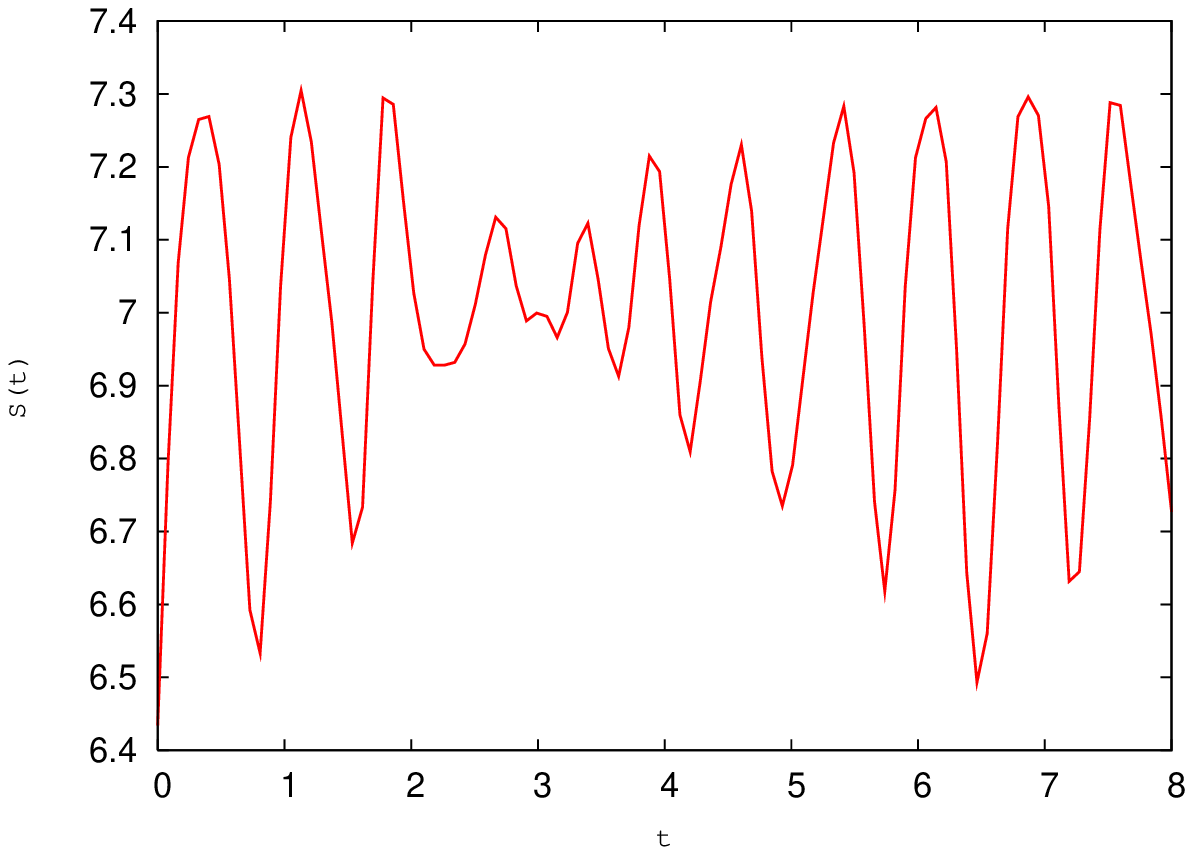}} & \\
          (c) $\omega= 2.0$ &\\
    \end{tabular}
  \end{center}
\caption{(color online) Plot of $S(t)$ with time $t$ for $g=100$ and various trap widths.}
\end{figure}
observe large amplitude oscillation and wavelength of oscillation increases as the first maximum occurs at $t=0.64$ for $\omega=1.0$ whereas the same occurs at $t=0.8$ for $\omega=0.85 \omega_0$. However for $\omega=2.0 \omega_0$ we observe the effect of inhomogenity and nonlinearity in the dynamics of $S(t)$.  In Fig.~8 we observe strong nonlinear and aperiodic fluctuating oscillation in $S(t)$ which indicates that the system is far away from linearity.

The results for $g=500$ in various trap sizes are presented in Fig.~9 where we observe very sharp change in the amplitude of oscillation in $S(t)$ for different trap sizes. For very tight trap as bosons become strongly inhomogeneous and highly correlated the effect of nonlinear interaction becomes strong and $S(t)$ exhibits strong aperiodic oscillation in $S(t)$. 
\begin{figure}[hbpt]
\begin{center}
    \begin{tabular}{cc}
      \resizebox{80mm}{!}{\includegraphics[angle=0]{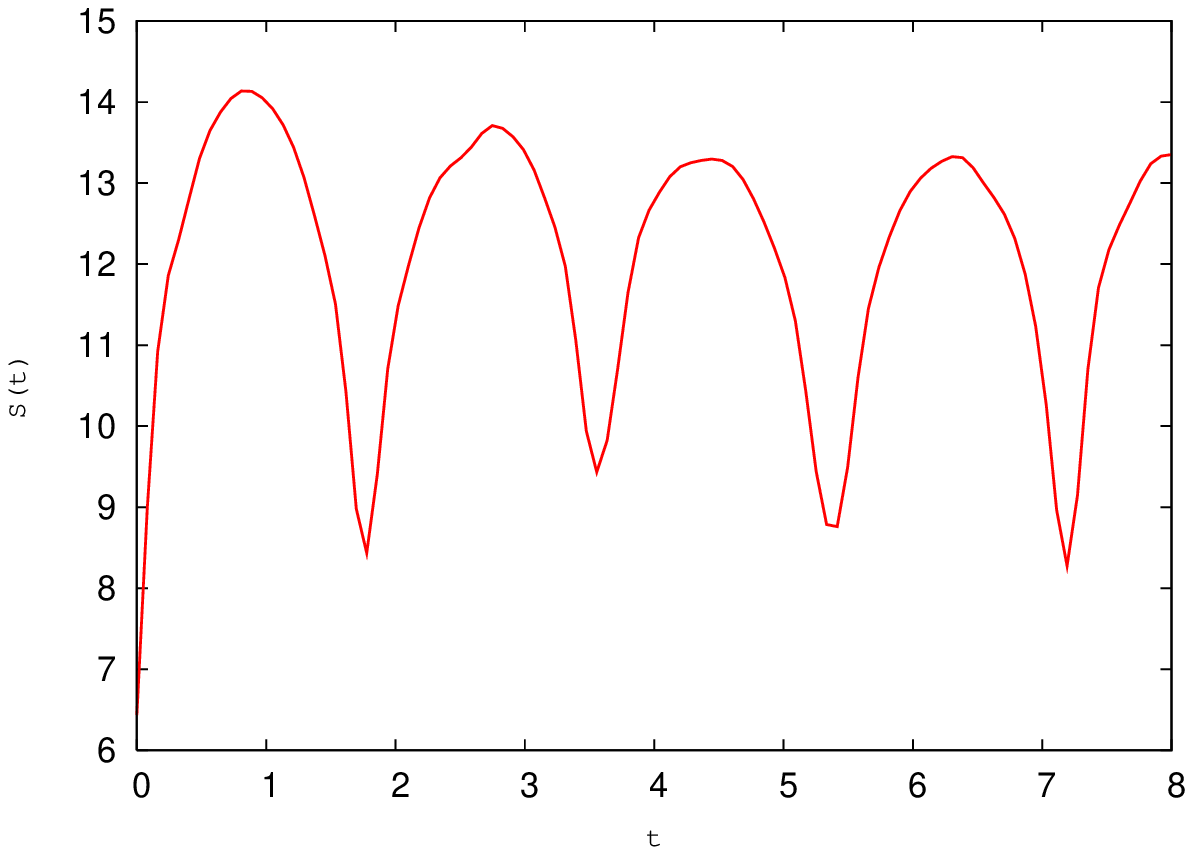}} &\\
         (a) $\omega =0.85$ &\\
      \resizebox{80mm}{!}{\includegraphics[angle=0]{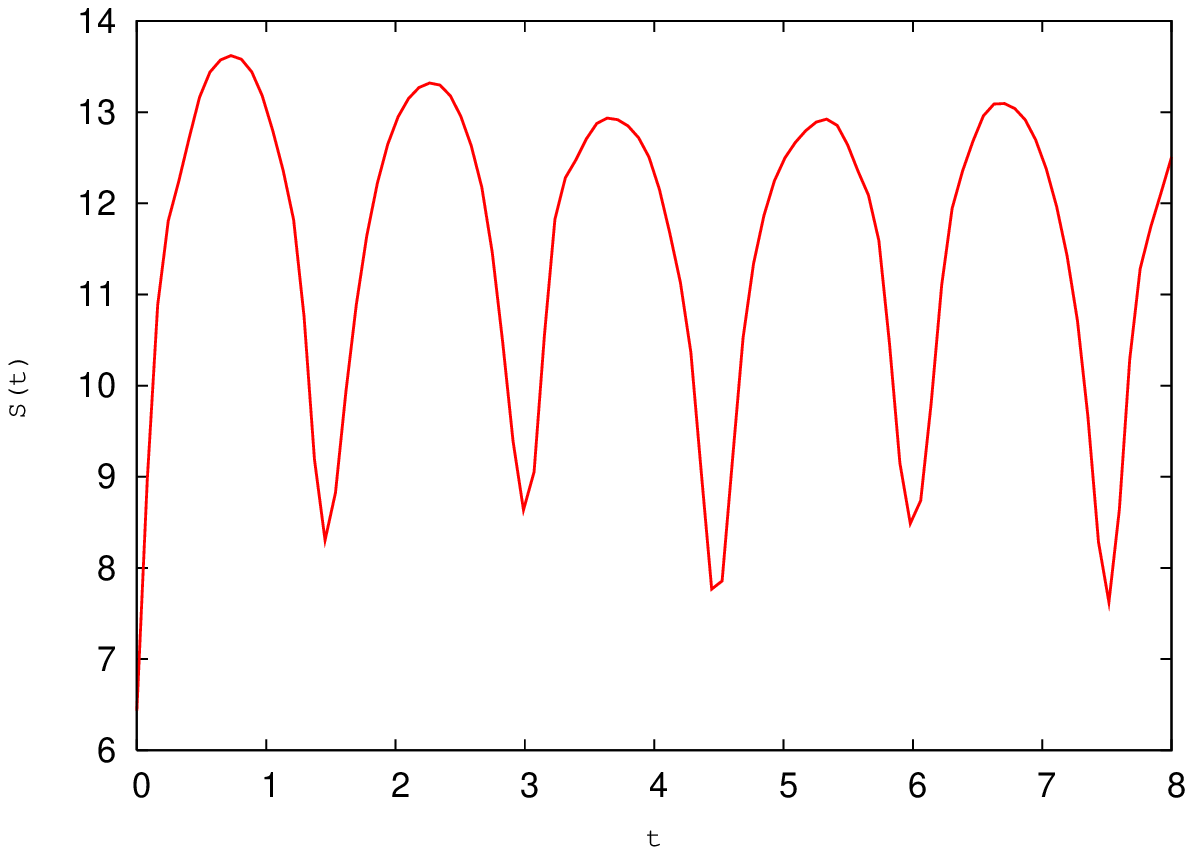}} & \\
         (b) $\omega =1.0$ &\\  
      \resizebox{80mm}{!}{\includegraphics[angle=0]{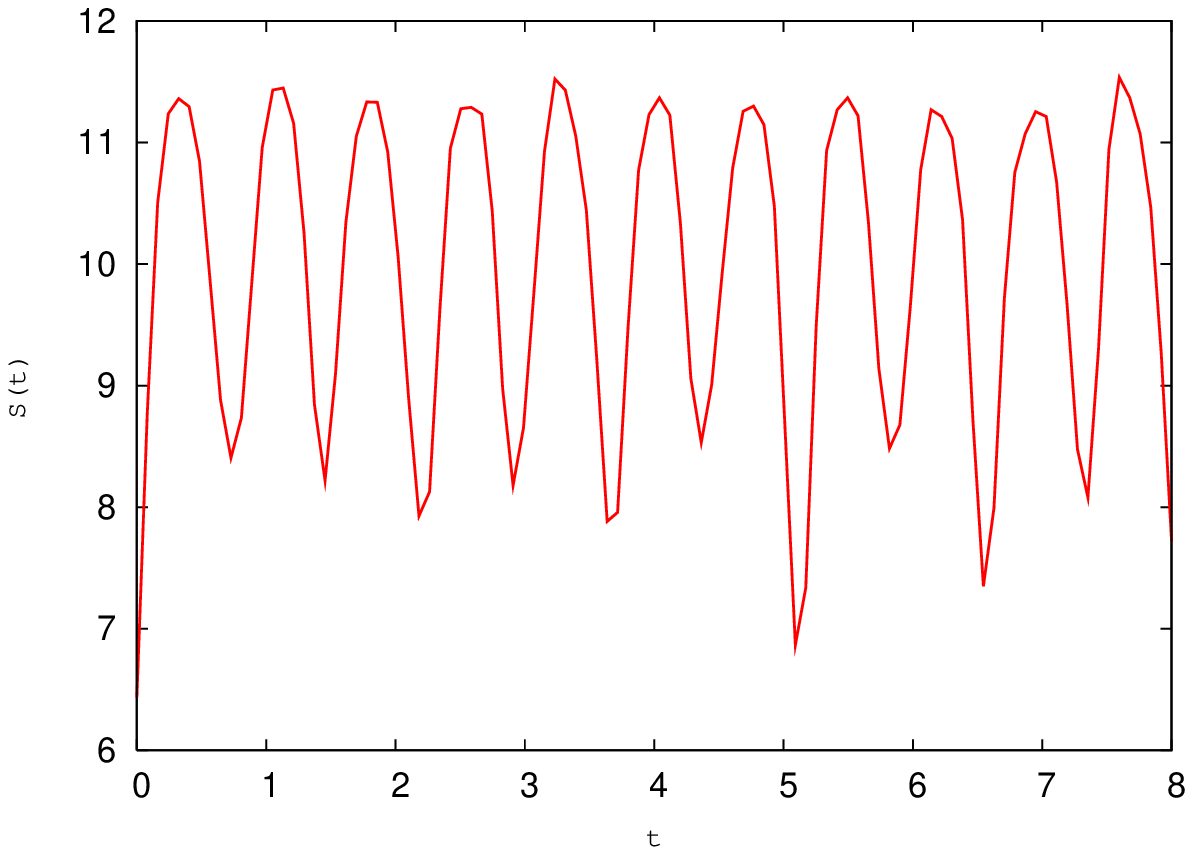}} & \\
          (c) $\omega= 2.0$ &\\
    \end{tabular}
  \end{center}
\caption{(color online) Plot of $S(t)$ with time $t$ for $g=500$ and various trap widths.}
\end{figure}

In ordinary thermodynamics, entropy is a measure of disorder in the system. However, defining disorder $\Delta$ as $\Delta=\frac{S}{S_{max}}$, where $S$ and $S_{max}$ refer to the same equation of state, Landsberg defined a order parameter $\Omega$~\cite{Landsberg} by $\Omega = 1 - {S}/{S_{max}}$. $\Omega$ = 1 corresponds 
to perfect order and $\Omega$ = 0 corresponds to randomness. It has been shown by a whole class of examples that the entropy $S$ and the order parameter $\Omega$ are two decoupled concepts~\cite{CP,Massen,Landsberg}. For the sake of completeness, we have also calculated the Landsberg's order parameter $\Omega$ at $t=0$. In Table $1$ we present the values of $\Omega$ for various nonlinearities.

\vskip 5pt
\begin{center}
\begin{table}[!h]
\caption{Landsberg's order parameter for various nonlinearity. }
\begin{center}
\begin{tabular} {|l|l|}
\hline
 $g$ &  $\Omega$ \\ \hline
10  &   $9.704 \times 10^{-5}$\\ \hline
100  &  $9.851 \times 10^{-5}$  \\ \hline
500 &   $1.961 \times 10^{-4}$ \\ \hline
\end{tabular}
\end{center}
\end{table}
\end{center}
\vskip 5pt

It implies that by increasing the number of particles, the system 
becomes more ordered. This observation agrees well with earlier observation
in atoms and clusters~\cite{CP, Massen}. This is also consistent with the
fact that the entropy and order are decoupled unlike the case in
thermodynamics~\cite{Landsberg}.

Lastly our calculated values can be correlated with the experimental observation. As stated earlier that the most relevant feature of trapped Bose gas is that they are inhomogeneous due to external trapping potential. The important consequense is BEC shows up not only in momentum space like noninteracting homogeneous ideal Bose gas, but also in coordinate space~\cite{Dalfovo}. This dual possibility of investigating the effects of trapped condensation and associated measurement of both the density and velocity distribution are very interesting and novel. The density distribution of the atoms in the trap as a function of coordinate is measured by the optical method~\cite{Bradley}. It also provides the measurement of size of the cloud. Whereas other kind of distribution is the velocity distribution in momentum space which is measured by time-of-flight measurements. The atoms were left to expand by switching off the confining trap and the velocity distribution is imaged. Thus the dual possibility to obtain the density distribution corresponds to the Fourier transformation of the condensate wave function in the momentum space. As the density in coordinate and momentum space are the basic ingredients to calculate the Shannon information entropy, our theoretical results can be correlated with the experimental observation. This is also nicely demonstrated in Fig. 2(a) and 2(b), where the opposite diverging behavior in $S_{r}$ and $S_{k}$ nicely demonstrate the experimentally observed critical point of stability.  Again the maximum value of information entropy can be accurately determined from Eq.~(11) where the size of the condensate $<r^2>_t$ and the kinetic energy of the system are known from the experimental results. Thus our theoretical results provide rich physics which can be tested in modern experiments.

Another relevant question is how to correlate the theoretical $T=0$ temperature results with those of experimentally measurable quantities which would be for $T>0$. In the real experimental situation the condensate temperature is $\sim$ nk as said earlier. Thus in the presence of thermal cloud, there is an interaction between the condensate and non-condensed atoms and thermal fluctuation also plays an important role. However in the present study we consider the $T \textgreater 0$ but $T \textless T_{c}$ ($T_c$ being the critical temperature) where there is a macroscopic population in the single lowest quantum state, i.e. the condensate is absolutely formed and there is no thermal fluctuation. Thus our system does not have thermally excited atoms. Apparently it may contradict the experimental situation, however in the presence of external trap as the interaction energy is very small compared to the trap energy, one can safely use the zero-temperature GP equation in the description of condensate. However when $T \textgreater T_{c}$, there will be thermal distribution of atoms in the higher states, the ground state is depleted and to study such system we must use temperature dependent formalism. Thus our $T=0$ temperature results are particularly valid in the length scales from the interparticle distance to the healing length which is the relevant range in current experiments where $na_{sc}^3 \ll 1$.

\section{Conclusion}        

Shannon entropies are generally used to examine the localisation 
of the distribution function in the position and momentum space.
In the present paper the dynamics of information entropies and 
the EUR are studied with varying interparticle potential of the
interacting trapped bosons and varying trap width. The system draws special attention
due to the presence of external trap. 
We observed that for weak interaction (i.e 
small $g$) the system shows periodic behavior irrespective of trap widths. However with increasing
interaction strength $g$, irregular behavior comes in and also the effect of varying trap width becomes more pronounced. We stress that 
since we choose the parameters like  $\omega_0$, $a_{sc}$ etc. as
those of JILA experiment and also restricted $N$ up to few thousands only, GP equation
is good enough to describe it. 
For more dense system with stronger interaction strength (i.e. larger $N$ and $a_{sc}$)
inter-atomic correlations become important. So some other theoretical technique 
incorporating correlations needs to be applied. Another important
observation is that the EUR is maintained throughout time for all cases
and thereby reaffirms its generality as an earlier study has verified it numerically
for different number of bosons in the trap.
The increase of $\Omega$ with $g$ (i.e. N, as we kept $a_{sc}$ fixed) is consistent
with earlier study and confirms again that the entropy and order are decoupled.
However the dynamic evolution of $\Omega$ with time for various $g$ remains to
be studied. 

Also another possible extension of our present work is to study the dipolar BEC. As mentioned earlier the interaction potential for dipolar bosons consists of two parts-the long-range anisotropic dipole-dipole interaction and the short-range $\delta$ potential characterized by the $s$ wave scattering length $a_{sc}$. In recent experiments $s$ wave interaction has been tuned keeping the strength of the dipole-dipole interaction fixed. It would be interesting to study its effect on the quantum properties of such dipolar BEC. Finally as described earlier, all our results are directly related to the experimentally measurable quantities and therefore they are experimentally accessible even with available set up.

We thank Prof. T. K. Das and Prof. S. E. Massen for some helpful discussions. 
Also Prof. Luca Salasnich is acknowledged for providing the numerical code
for solving GP equation and for his kind suggestions for preparing the manuscript.
This work has been partially supported by FAPESP (Brazil), 
Department of Science and Technology (DST, India) and 
Department of Atomic Energy (DAE, India). 
B.C. wishes to thank FAPESP (Brazil) for providing financial 
assistance for her visit to the Universidade de S\~ao Paulo, Brazil, 
where part of this work was done. SKH acknowledges the 
Council of Scientific and Industrial Research (CSIR), India for the Senior 
Research Fellowship SRF (NET).

\end{document}